\documentclass{elsart}

\usepackage{epsfig}

\usepackage{amssymb}

\begin{document}

\begin{frontmatter}



\title{Study of Scintillator Strip with Wavelength Shifting Fiber and Silicon Photomultiplier.}


\author{V.Balagura   \thanksref{ITEP}}
\author{M.Danilov    \thanksref{ITEP}}
\author{B.Dolgoshein \thanksref{MIPHI}}
\author{S.Klemin     \thanksref{MIPHI}}
\author{R.Mizuk      \thanksref{ITEP}}
\author{ P.Pakhlov   \thanksref{ITEP}}
\author{E.Popova     \thanksref{MIPHI}}
\author{V.Rusinov    \thanksref{ITEP}}
\author{E.Tarkovsky  \thanksref{ITEP}}
\author{I.Tikhomirov \thanksref{ITEP}}

\address[ITEP]{Institute for Theoretical and Experimental Physics, B.Cheremushkinskaya 25, Moscow, 117259, Russia}
\address[MIPHI]{Moscow Engineering and Physics Institute, Kashirskoe sh. 31, Moscow, 115409, Russia}

\begin{abstract}
The performance of the $200\times2.5\times1$ cm$^3$ plastic scintillator strip 
with wavelength shifting fiber read-out by two novel photodetectors called
Silicon PhotoMultipliers (SiPMs) is discussed. The advantages of SiPM relative to
the traditional multichannel photomultiplier are shown.
Light yield and light attenuation measurements are presented.
This technique can be used in muon or calorimeter systems.

\end{abstract}

\begin{keyword}
Scintillation detectors \sep wavelength shifting fibers \sep silicon photomultiplier

\PACS 29.40Mc \sep 29.40Vj
\end{keyword}
\end{frontmatter}

The detection of charged particles with plastic scintillators, wavelength shifting (WLS) fibers
and multichannel photomultipliers 
is a well known, efficient and robust technique (see, e.g.~\cite{MINOS}). 
However it has severe limitations. First, photomultipliers can not work in the magnetic field.
For scintillator counters inside a magnet one should bring the light
out by clear fibers. This complexifies the detector and leads to some light losses. 
Second, fibers from different scintillator counters should be assembled together in a bundle
attached to the multichannel photomultiplier. This is not always easy to arrange. 
Finally, calibration and monitoring of a multichannel photomultiplier is not a simple task.

In this work performed at ITEP (Moscow) we use the novel photodetector called Silicon Photomultiplier (SiPM)~\cite{SiPM1,SiPM2}
instead of the traditional photomultiplier.
It is the matrix of 1024 = 32$\times$32 independent silicon photodiodes covering the area of 1$\times$1~mm$^2$.
Each diode has its own quenching polysilicon resistor of the order of a few hundred k$\Omega$. All diode-resistor pairs called pixels 
later on are connected in parallel.
A common reverse bias $V_{bias}$ is applied across them.
Its magnitude of the order of 40--60 V is high enough to start the Geiger discharge if any free charge carrier appears in the $p-n$ junction 
depletion region. 
The diode discharge current 
causes a voltage drop across the resistor. This reduces the voltage across the diode below the breakdown voltage $V_{breakdown}$ and
the avalanche dies out. One diode signal is 
$Q_{pixel}=C_{pixel}(V_{bias}-V_{breakdown})$ where $C_{pixel}$ is the pixel capacitance. Typically $C_{pixel}\sim 50$~fF and
$\Delta V = V_{bias}-V_{breakdown}\sim 3 V$ yielding $Q_{pixel}\sim 10^6$ electrons. 
Such an amplification is similar to the one of a typical photomultiplier and 3--4 orders of magnitude larger than the 
amplification of an Avalanche Photo Diode (APD) working in the proportional mode. 

\begin{figure}[htbp]
\begin{center}
\epsfig{figure=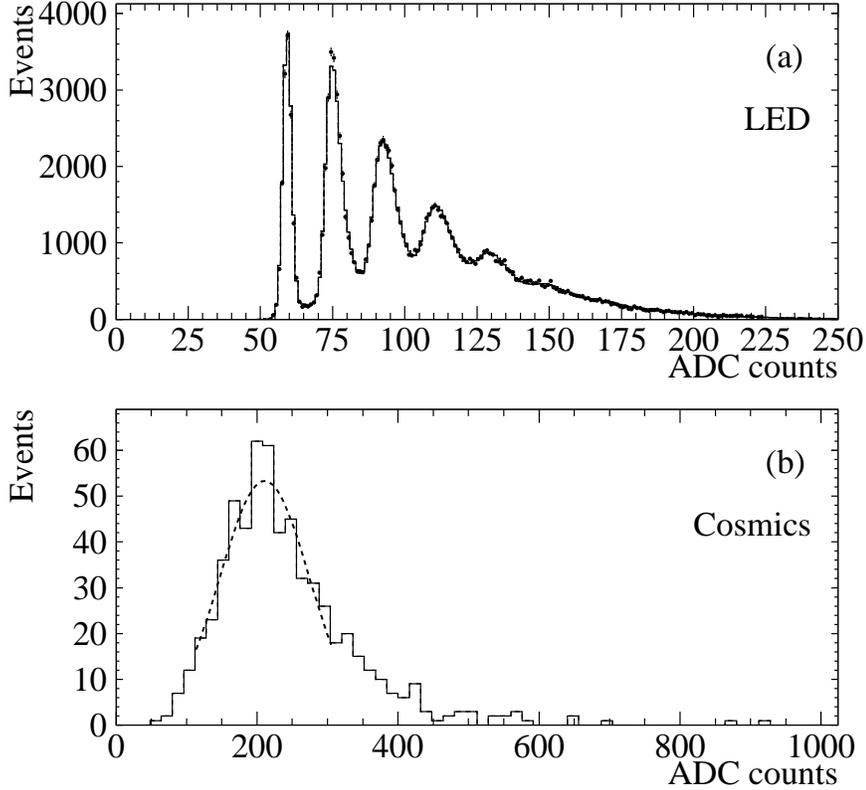,width=12cm,angle=0}
\end{center}
\caption{\em SiPM signal spectra.
 (a): SiPM is illuminated by short weak flashes produced by LED. The fit curve is described in detail
      in Appendix~\ref{Fit}.
 (b): Typical SiPM response to cosmic particles. 
The dashed line shows the fit of the region around the peak to the Gaussian distribution.}
\label{ADC}
\end{figure}

$Q_{pixel}$ does not depend on the number of primary carriers which start the Geiger discharge. 
Thus each diode detects the carriers created e.g. by a photon, a charged particle or by a thermal noise with the same response 
signal of $\sim 10^6$ electrons. Moreover the characteristics of different diodes inside the SiPM are also very similar. When fired, they produce 
approximately the same signals. 
This is illustrated in Fig.~\ref{ADC}a. It shows the SiPM response spectrum when it is illuminated by weak flashes 
produced by 
a Light Emitting Diode (LED). 
The spectrum is obtained by integrating the SiPM signal during 120~nsec and using analog-to-digital converter (ADC). 
The integration time is big enough to contain most of the SiPM signal which
lasts a few tenths of nsec. 
First peak in this figure is the pedestal. 
The second one is the SiPM response when it detects exactly one photon. It is not known which diode inside the SiPM
produces the signal
since all of them are connected to the same output. 
However since the responses of all pixels are similar, 
the peak width is small.
If several pixels in the SiPM are fired, the net charge signal is the sum of all charges.
The third, forth and so on peaks in Fig.~\ref{ADC}a correspond to 2, 3, ... fired pixels. 
Note that the peaks are equidistant.

If $n$ pixels are fired the corresponding peak width is approximately 
$ \sigma_n = \sqrt{n\sigma^2+\sigma_{pedestal}^2}, $
where 
$\sigma_{pedestal}$ is the pedestal width and $\sigma$ describes
the spread of signals $Q_{pixel}^i$ from different pixels.
Typical value of $\sigma$ divided by the average signal of one pixel is
 $\frac{\sigma}{<\!\!Q_{pixel}^i\!\!>}\approx 15\%$.
For large $n$ the peaks start to overlap and the spectrum loses its peak-like structure. This is illustrated in Fig.~\ref{ADC}b which 
shows the SiPM response to the photons created by the minimum ionizing particle (MIP) 
in the scintillator detector.
In this case the average number of fired pixels is 8.5. 
This is still much smaller than the total number of pixels in the SiPM.
Therefore the signal is approximately proportional to the number of photons and to the $dE/dx$ losses in the scintillator. 
For much larger $n$ the saturation effects come into play, and the whole dynamic range is limited by the finite number of SiPM pixels (1024 in our 
case~\footnote{SiPMs can be produced with different number of pixels in the range 500--4000.}). 
The SiPM photodetection efficiency depends on the light wave length and the overvoltage $\Delta V$.
Typical value is about 10--15\% for the green light. It includes geometrical inefficiency due to dead regions in the SiPM between the pixels. 

Thus SiPM and traditional photomultipliers have similar gain and efficiency.
However, SiPM is approximately twice cheaper than one channel in the multichannel photomultiplier.
In addition it can work in the magnetic field, so there is no need in the light transportation out of the magnetic field with clear fibers.
The SiPM size matches well with the size of WLS fiber. SiPM can be mounted directly on the detector which simplifies the detector design.
SiPM has a quite high noise rate of about 2~MHz at 0.1 photoelectron threshold. However the noise rate drops fast with increasing threshold. 
This will be discussed in more detail later. 

\begin{figure}[htbp]
\begin{center}
\epsfig{figure=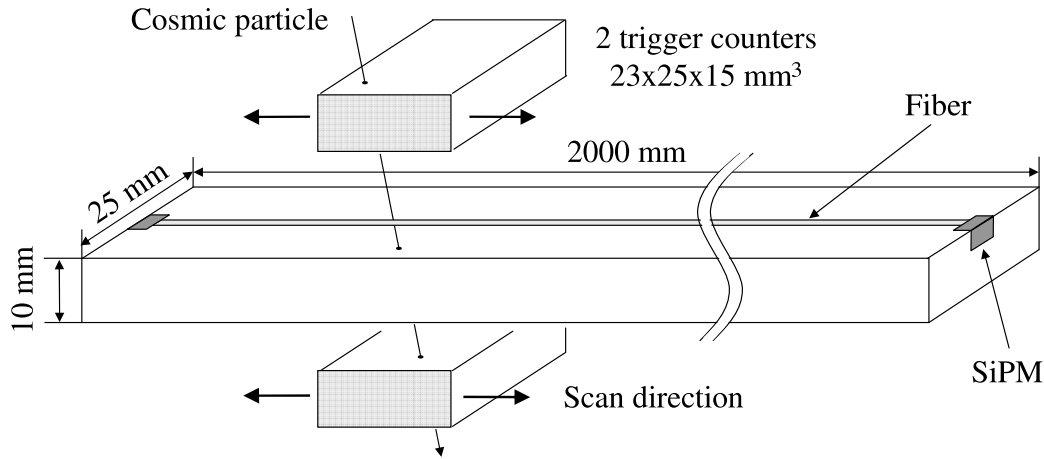,width=6cm,angle=270}
\end{center}
\caption{\em The layout of the test bench (not in scale).}
\label{setup}
\end{figure}

Our detector consists of a $200\times2.5\times1$ cm$^3$ plastic scintillator strip and a wavelength shifting fiber 
read-out by two SiPMs installed at the strip ends (see Fig.~\ref{setup}). 
The scintillator strip is produced at the ``Uniplast'' enterprise in Vladimir, Russia. This is one of the biggest 
plastic scintillator producers 
in the world. The scintillator
for the electromagnetic calorimeters of PHENIX, HERA-B and LHC-B experiments has been produced there. 
The strip is extruded 
from the granulated polystyrene with two dyes (1.5\% of PTP and 0.01\% of POPOP). 
The Kuraray multiclading WLS fiber Y11 (200) with 1~mm diameter is put in the 2.5~mm deep groove in the middle of the strip.
It is fixed there only by the friction. No gluing or any other kind of optical coupling is used to attach the fiber to the strip or to the SiPM.
The SiPMs are mounted to the special slots made in the scintillator and are also kept there only by the friction. 
There is about 200~$\mu$m air gap between 
the fiber end and the SiPM. We did not study the light output dependence on the size of the gap. 200~$\mu$m is chosen from the mechanical point of view 
as the 
minimal gap which ensures that the fiber can not scratch the sensitive SiPM surface.
To improve the light collection efficiency, the strip is wrapped in the Superradiant VN2000 foil produced by the 3M company. 

We use the cosmic particle trigger consisting of a pair of
2.3$\times$2.5$\times$1.5 cm$^3$ scintillator counters 
placed above and below the strip (see Fig.~\ref{setup}). The SiPM spectra like the one 
shown in Fig.~\ref{ADC}b, are obtained
for different positions of the trigger counters along the strip. 
The SiPM signal is first amplified by the preamplifier with the gain of about 100. The preamplifier nonlinearity 
is $<0.3\%$. Then standard analog-to-digital converter ADC LeCroy 2249A is used with the sensitivity
of 0.25~pC/channel. The charge integration time is 120~nsec.

In parallel to collecting the cosmic ray data, the strip is illuminated periodically by the flashes produced by
the blue LED. The LED light is transported to the strip through the hole in the VN2000 foil by a clear fiber. 
The duration of the flash is $\sim$20~nsec. 
The corresponding SiPM response spectrum shown in Fig.~\ref{ADC}a has been already discussed. 
It is used for the SiPM calibration. 
As a first approximation one can take the distance between adjacent peaks in the spectrum as the measure of the average signal of one pixel 
\mbox{$<\!\!Q_{pixel}^i\!\!>$}.
Using this value one can always convert a given number of ADC counts to the number of fired SiPM pixels. 
In practice to make the calibration more accurate, the value of \mbox{$<\!\!Q_{pixel}^i\!\!>$} is obtained from the fit described in detail in Appendix~\ref{Fit}.
The fitting function for the SiPM response spectrum is constructed in the following way.
 First, the function parametrizing the distribution of one pixel signals $Q_{pixel}^i$
is chosen. We assume either normal or so called log-normal shape (see Appendix~\ref{Fit}). 
The latter has one extra parameter which allows
to describe possible asymmetry between left and right parts of the $Q_{pixel}^i$ distribution. 
It gives better description of experimental data. 
The symmetric normal shape is used for comparison to estimate systematic errors. 
The signals corresponding to firing of two, three etc. 
pixels are modelled using double, triple etc. convolutions of one pixel distribution. Then all such
distributions are summed up with the weights equal to
the probabilities $p_N$ to fire exactly $N=0, 1, 2, 3, \ldots$ pixels. 

The probabilities $p_N$ are determined in the following way. 
It is assumed that the number of {\it photons} detected by SiPM follows the Poisson law. 
To relate the number of fired {\it pixels} ($N$) and the number of detected {\it photons} ($n$)
one should take into account the so called SiPM interpixel cross-talk. 
This is a rather well known effect (see e.g.~\cite{SiPM2}).
The Geiger discharge of one diode inside the SiPM can fire with some probability another diode and so on. 
Therefore in general $N$ can be larger than $n$. 
Every photon can fire $k=1, 2, 3, \ldots$ pixels and 
the corresponding probability drops approximately exponentially, i.e. as $\epsilon^k$ with some constant $\epsilon$.  
The parameter $\epsilon$ increases with the applied voltage. 
Thus the interpixel cross-talk shifts the $p_N$ distribution towards the larger values. 
The parameter $\epsilon$ can be estimated from the observed deviation of $p_N$ probabilities
from the Poisson distribution. Knowing $\epsilon$ one can calculate the average number of 
pixels fired by one initial photoelectron which is equal to $1/(1-\epsilon)$ (see Appendix~\ref{Fit}). 
From the best fit with the asymmetric 
log-normal shape of one pixel signals $Q_{pixel}^i$ this value is found to be
$1.22\pm 0.05$ for the left SiPM and $1.10\pm 0.04$ for the right one. 
The errors here are obtained by comparing the fit results of the calibration spectra collected for the same SiPM 
in different time periods. They reflect the uncertainty of cross-talk determination in the fit procedure.

The resulting fitting function for the SiPM response spectrum
thus depends on the shape of one pixel signal, average number of detected photons \mbox{$<\!\!n\!\!>$} and the
parameter $\epsilon$. In the end it is additionally convoluted with the ``pedestal'' spectrum collected 
with the random trigger and with the same gate of 120~nsec. 
It is not shown here since it essentially contains only pedestal and a small fraction of entries 
near the one pixel peak. Convolution with this spectrum
models the contribution of random noise hits not synchronised with the LED flash.
A convenient fast algorithm of calculating the described SiPM spectrum is given in Appendix~\ref{Fit}.

It is found from the fit that the peaks in the calibration spectra are slightly nonequidistant. 
This is probably due to 
nonlinearity in ADC. Therefore in the final fit with the asymmetric log-normal shape of the $Q_{pixel}^i$ distribution,
the positions of the first 5 peaks 
are allowed to float from their nominal positions. 
The measured differences in spacing between the peaks in the same spectrum 
are used to estimate the systematic error of the average signal \mbox{$<\!\!Q_{pixel}^i\!\!>$}.
For 14 measurements presented in the following these errors vary between 0.3 and 1.1 ADC channels.
The resulting $\chi^2$ per degree of freedom of the fit is found to be in the range $1.1-2.4$. 
The average number of entries in one calibration spectrum is 70 000.
In the alternative fit with the normal shape of one pixel signal the peaks are required to be equidistant. 
In this case the obtained 
$\chi^2$ is significantly worse: $3-11$. 
Such a fit is used only for the comparison.

The fit with the asymmetric log-normal shape shows that one pixel signal distribution has smaller left and bigger 
right tail. This means that the mean value of this distribution \mbox{$<\!\!Q_{pixel}^i\!\!>$} is slightly bigger than the position of the maximum,
and is bigger than the spacing between the peak maxima in 
Fig.~\ref{ADC}a. For the normal shape this effect is absent since 
the maximum and the average coincide. The determined value of \mbox{$<\!\!Q_{pixel}^i\!\!>$} 
in this case is found to be 
$\sim 5\%$
smaller. Another difference between the two fits is the measured value of the cross-talk parameter $1/(1-\epsilon)$. For the normal shape it is found to
be 17\% larger (1.43 for the left SiPM and 1.28 for the right one). 
Thus in total the fit with the symmetric shape of the single pixel peak gives 12\% 
bigger signal produced by one detected photon \mbox{$<\!\!Q_{pixel}^i\!\!>/(1-\epsilon)$}.

Knowing \mbox{$<\!\!Q_{pixel}^i\!\!>/(1-\epsilon)$} one can calculate the light yield of the detector 
for the minimum ionizing particle from the spectra like the one shown in Fig.~\ref{ADC}b.
To be conservative in our estimates,
we take the maximum of the Landau distribution instead of the average value, 
subtract the pedestal position and divide it by  \mbox{$<\!\!Q_{pixel}^i\!\!>/(1-\epsilon)$} to
get the number of photons. 
Note that if one uses the mean value of the spectrum in Fig.~\ref{ADC}b instead of the maximum position, 
the estimated number of photons increases by about a factor of 1.1 due to 
Landau tail. 
To determine the maximum position, the 
region around the Landau peak is fit to the Gaussian distribution as shown in Fig.~\ref{ADC}b. 
Varying the fit region results in some systematic uncertainty in the distance between the maximum 
and the pedestal position at the level of 1--3\%.

Due to finite sizes of the trigger counters the selected cosmic particles are not strictly vertical.
Thus they produce slightly more scintillation light than the minimum ionizing particle at normal incidence to the strip.
To correct for this effect, a simple simulation is made. It assumes that the angular distribution of all cosmic particles
coincides with the one of cosmic muons and has the form $\sim\cos^2 \theta$ where $\theta$ is the angle with respect to the vertical 
direction~\cite{PDG}.
In this way it is found that in average the path length of triggered particles inside the scintillator is 10\% 
larger than the strip thickness. Assuming that the light yield is also 10\% larger, 
the measured average numbers of photons
are divided by 1.1.

\begin{figure}[htbp]
\begin{center}
\epsfig{figure=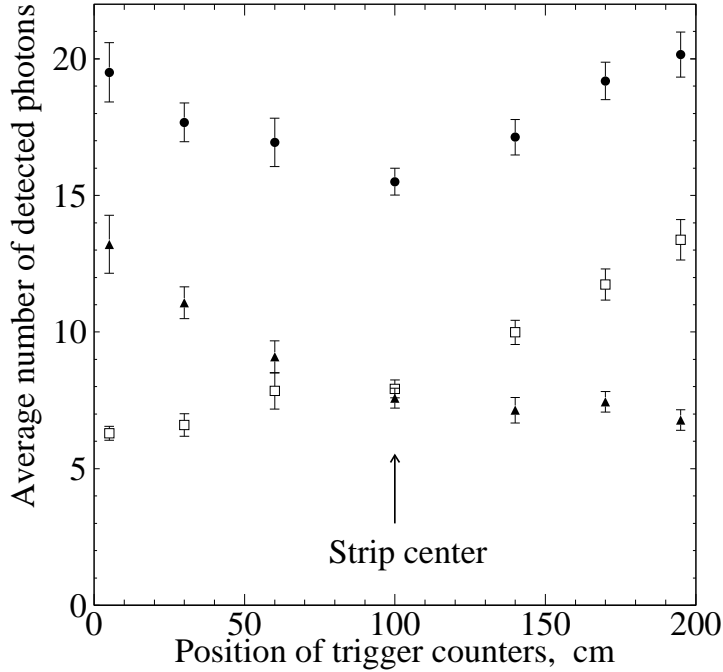,width=10cm,angle=0}
\end{center}
\caption{\em Average number of photons detected by the left (triangles) and right (squares) SiPMs for normally incident cosmic 
ray particle versus the position of the trigger counters along the strip. 
The upper curve (filled circles) is the sum of the two SiPM signals.}
\label{res}
\end{figure}

The resulting light yield is shown in Fig.~\ref{res} for different positions of the trigger counters along the strip.
The errors include contributions from the uncertainties in the position of the maximum in Landau 
spectra (Fig.~\ref{ADC}b) estimated by varying the fit intervals (1.6\% in average),
errors of \mbox{$<\!\!Q_{pixel}^i\!\!>$} due to nonlinearity of ADC (3\% in average) and errors of the
cross-talk parameter $1/(1-\epsilon)$ estimated from the assumption that it should be stable in time (4\%).
The difference in the value of \mbox{$<\!\!Q_{pixel}^i\!\!>/(1-\epsilon)$} between the two alternative fits (12\%) is not included here as an error.
Filled triangles and open squares denote the left and right SiPM respectively. 
One can see that for the 2~m strip detector the
transportation of light from the far end attenuates it by about a factor of 2.
The upper filled circles in Fig.~\ref{res} show the sum of the two SiPM signals.
For the minimum ionizing particle at normal incidence 
the measured average number of detected photons varies in the range 15.5 -- 20.2 depending on the position along the strip.
The difference in the light yields from the center and the end of the strip can be corrected if 
the position along the strip is known. Such a correction at the level of $\pm13\%$ can significantly improve the uniformity of the detector response 
for precise calorimetric measurements. 
%
%

In the worst case when the particle passes through the strip center there are 
15.5 detected photons. As it was explained above the alternative fit of the calibration spectrum with symmetric normal shape of one pixel signal 
and equidistant peaks produces bigger value of \mbox{$<\!\!Q_{pixel}^i\!\!>/(1-\epsilon)$} and correspondingly 
smaller number of detected photons. For the center
of the strip it gives 13.7 detected photons. 
In spite of the fact that the alternative fit gives worse description of the calibration spectra and
significantly larger $\chi^2$, we take conservatively this number for the following estimations.
In case of Poisson statistics it corresponds to 98\% efficiency at the threshold $\ge 7$ photons.
Such a high threshold is needed to reduce the SiPM noise rate. 
To express the same requirement in terms of fired pixels a simple modelling is made. 
Instead of the sum of two SiPM signals we simulated the response of one SiPM to 13.7 Poisson distributed photons.
For consistency we took the model with symmetric normal shape of one pixel signal. Its width $\sigma$
and the cross-talk
are taken as the average of the corresponding values of two SiPMs.
Since the SiPMs have different gains, $\sigma$ is normalized to the
distance between adjacent peaks before averaging.
The left part of the modelled spectrum shows that 98\% efficiency corresponds to the requirement to have $\ge 8$ fired pixels.
With this requirement the efficiency averaged over the whole strip exceeds 99\%. 

\begin{figure}[htbp]
\begin{center}
\epsfig{figure=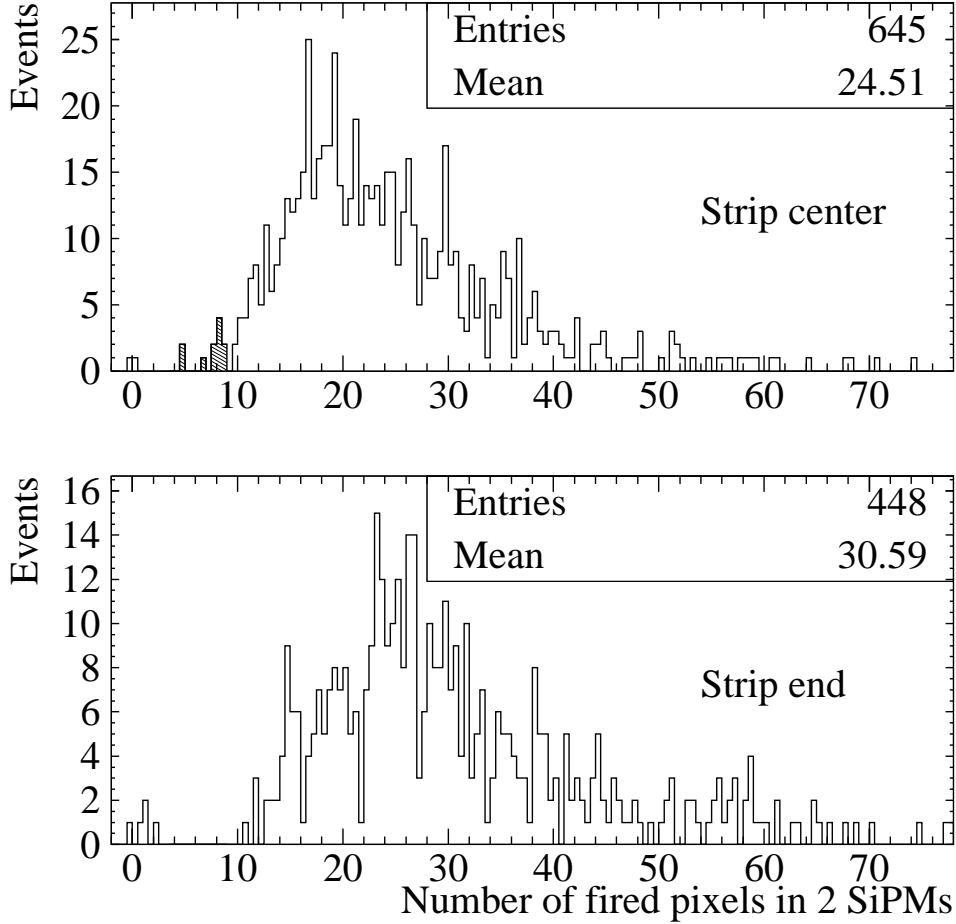,width=14cm,angle=0}
\end{center}
\caption{\em Number of fired pixels in two SiPMs when trigger counters are located at the strip center (upper plot) and at the ends (lower plot).}
\label{npixels}
\end{figure}

This estimate can be checked with the data.
It is convenient to express the SiPM response signals in terms of fired pixels. This is achieved simply by shifting and scaling the ADC counts.
Position of zero and the scaling factor are determined from the calibration spectrum like in Fig.~\ref{ADC}a.
The distribution of the number of pixels fired in both SiPMs by a cosmic particle is shown
in Fig.~\ref{npixels}. 
The top and bottom spectra correspond to two extreme cases when the particle passes through the center or the end of the strip.
A few events around zero belong to the pedestal. It appears
here due to imperfectness of the trigger.  The plots are not corrected
for the factor 1.1 which was introduced above to take into account not normal incidence of cosmic particles.
Therefore to estimate the inefficiency of the requirement to have $\ge 8$ fired pixels for normal incidence one should  
count the entries between the pedestal and the value $8\cdot 1.1 \approx 9$.  This
part of the spectrum is hatched. There are 11 such events in the top
plot. They correspond to the $1.7\pm0.5$\% inefficiency which agrees with the
calculations above. The inefficiency averaged over all positions of the trigger counters along the strip is found to be $0.7\pm0.2$\%. 

In the practical use of SiPM the complicated fit of calibration spectra described above is not 
necessarily required.
The minimum ionizing particle spectrum will be measured directly. The MIP peak position 
will be used for the absolute calibration of the detector.
For example this was done in~\cite{SiPM2} to calibrate the high granularity calorimeter prototype
consisted of about 100 scintillator tile detectors read out via WLS fiber and SiPM. 
Measuring the spacing between the peaks in the SiPM calibration spectra will be needed only to check
the time stability of the SiPM response and to apply corrections.

\begin{figure}[htbp]
\begin{center}
\epsfig{figure=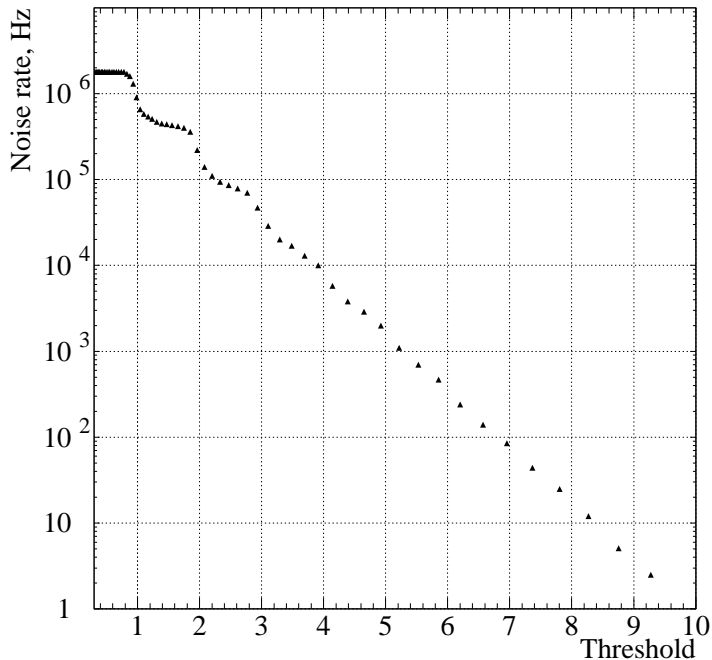,width=10cm,angle=0}
\end{center}
\caption{\em Typical SiPM noise rate versus the threshold expressed in the units corresponding to one pixel signals.}
\label{noise}
\end{figure}

The typical SiPM noise rate at a room temperature is shown in Fig.~\ref{noise}. It is measured without any trigger by counting the
number of SiPM signals higher than a given threshold. The rate starts at about 2~MHz.
The threshold value is expressed in the units corresponding to one pixel signals. Therefore one can see clear steps
at 1, 2, 3, 4 in the beginning of the plot. They 
correspond to the peaks in Fig.~\ref{ADC}a. 
For larger signals the step-like structure becomes smeared and the rate drops approximately exponentially with the threshold.
Since the SiPM signal is very short ($\lesssim 20$~nsec) the probability that two independent noise signals overlap in time is small.
Therefore big noise signals can be produced only by simultaneous correlated firing of several pixels caused by the interpixel cross-talk. 
Thus the exponential behaviour confirms the cross-talk model based on the geometric progression $1,\epsilon, \epsilon^2,\ldots$\ .
The exponential slope in Fig.~\ref{noise} is determined by the cross-talk probability $\epsilon$.

The rate of noise signals from two SiPMs with the 
threshold of 8 fired pixels can not be directly read off the plot in Fig.~\ref{noise}.
It depends on the type of the electronics which detects the coincidence between the SiPMs and calculates the total signal.
As an example one can assume that the threshold is set for the sum of two SiPM signals integrated 
during the same 120~nsec gate which is used in obtaining the amplitude spectra in Fig.~\ref{npixels}.
Comparing the gate with the 2~MHz SiPM noise rate one can see that there is a sizeable probability that two or even more noise signals 
can contribute to the net signal. This changes the exponential slope in Fig.~\ref{noise}.
%
%
The probability to get $\ge 8$ fired pixels in the random 120~nsec window is found to be $7\cdot 10^{-4}$.
It is measured with the random trigger. 
Clearly, the noise can be suppressed even further if the electronics utilizes the fact that two SiPM signals caused by real particles 
are closer in time than 120~nsec. 

In conclusion, the detector consisting of the $200\times2.5\times1$ cm$^3$ plastic scintillator strip,
the wavelength shifting fiber and two novel photodetectors called 
Silicon PhotoMultipliers has been constructed and tested.
This new technique can be used in the muon systems or in calorimeters.
For example it can be used in the muon system of the future International Linear Collider detector.
The tested scintillator detector has higher efficiency and by far higher rate capabilities than resistive plate 
chambers which are often used in muon systems.
SiPM has similar gain and efficiency as the traditional multichannel photomultiplier. It also has several advantages.
There is no need to use clear fibers to bring the light out of the magnetic field and
to arrange many fibers in one bundle attached to the multichannel photomultiplier. 
SiPM can be mounted directly on the strip end.
Its gain can be determined easily by observing the peaks corresponding to different number of fired SiPM pixels
(see Fig.~\ref{ADC}a). Finally it is approximately twice cheaper than one channel in a multichannel photomultiplier.
Further cost reductions are expected in case of a mass production.
The light yield and light attenuation measurements for the tested scintillator strip detector are shown in Fig.~\ref{res}.
The light yield of more than 15 detected photons per cosmic ray particle at normal incidence is obtained.
The light collection efficiency can be further increased by gluing the WLS fiber to the strip~\cite{gluing}. We plan to study this
possibility systematically in the future. 

 \appendix
\section{The fit procedure of the SiPM calibration spectrum}
\label{Fit}

To calibrate the SiPM, it is illuminated by the flashes produced by LED.
A typical SiPM response spectrum is shown in Fig.~\ref{ADC}a. 
This histogram is fit to the convolution ($R$) of the pedestal spectrum ($B$)
obtained with the random trigger when the 
LED is off and the SiPM response function to the LED flash ($L$) which will be described later. 
To model the SiPM response it is very convenient to use the Fourier transform.
It will be denoted in the following by $^{\mathrm{F}}$ superscript.
The Fourier transform of the convolution of $B$ and $L$ is the product
$R^{\mathrm{F}}=B^{\mathrm{F}} L^{\mathrm{F}}$. Assuming the stability of LED and the pure Poisson distribution of 
the photons detected by 
SiPM, $L^{\mathrm{F}}$ can be modelled as
$$ L^{\mathrm{F}} = \sum_{n=0}^{+\infty} \frac{e^{-\mu}\mu^n}{n!}(P^{\mathrm{F}})^n = \exp \{\mu(P^{\mathrm{F}} - 1)\},$$
where $P^{\mathrm{F}}$ is the Fourier transform of the response to exactly {\em one photon}, 
$\mu$ is the
average number of photons detected by the SiPM. We use the fact that the response to
$n$ photons is $n$ convolutions of $P$ and thus has a Fourier transform $(P^{\mathrm{F}})^n$.
Due to the interpixel cross-talk one photon can fire more than one pixel. To describe this effect
we approximate $P^{\mathrm{F}}$ by 
$$ P^{\mathrm{F}} = \frac{Q^{\mathrm{F}} + \epsilon (Q^{\mathrm{F}})^2 + \epsilon^2 (Q^{\mathrm{F}})^3 + \ldots + \epsilon^{k-1} (Q^{\mathrm{F}})^k
+ \ldots}{1 + \epsilon + \epsilon^2 + \ldots + \epsilon^{k-1} + \ldots} =
Q^{\mathrm{F}}\frac{1-\epsilon}{1-\epsilon Q^{\mathrm{F}}},$$
where $\epsilon$ describes the cross-talk probability, 
$Q^{\mathrm{F}}$ is the Fourier transform of the SiPM signal distribution when exactly
{\it one\/} random {\it pixel\/} is fired. 
The probability that one initial photon fires $k$ pixels drops proportionally to $\epsilon^{k-1}$. This is
described by the term $\epsilon^{k-1} (Q^{\mathrm{F}})^k$ in the enumerator. Denominator normalizes the
$P$ distribution to one.
The average number of pixels fired by one photon is 
$\frac{1+2\epsilon+3\epsilon^2+\ldots+k\epsilon^{k-1}+\ldots}{1+\epsilon+\epsilon^2+\ldots}=$
$(1-\epsilon)\cdot\frac{d(\epsilon+\epsilon^2+\epsilon^3+\ldots)}{d\epsilon}=$$1/(1-\epsilon)$.
As an approximation of $Q$ the normal or log-normal distribution is taken.
The former depends on two parameters: the width $\sigma$ and the position of the maximum $\Delta$.
The latter is the distribution of the variable whose logarithm is normally distributed:
$Q_{\rm log-n}(x, \sigma, \Delta, \eta) = 
\frac{\eta}{\sqrt{2\pi}\sigma\sigma_0}\exp(-\frac{\ln^2(1-\eta(x-\Delta)/\sigma)}{2\sigma_0^2}-\frac{\sigma_0^2}{2})$,
where $\sigma_0 = \frac{1}{\sqrt{2\ln 2}}\sinh^{-1}(\sqrt{2\ln 2}\eta)$. 
$\Delta$ is again position of the maximum
and $\eta$ is the additional asymmetry parameter. 
Due to asymmetry, $\Delta$ can be different from the average signal of one pixel, i.e. the mean of the $Q$
distribution \mbox{$<\!\!x_Q\!\!>\ne \Delta$}.

If $Q$, $B$ and $R$ functions are normalized so that they have unit integrals, the resulting
formula for the Fourier transform of the fit function is 
$$ N \cdot R^{\mathrm{F}} = N \cdot B^{\mathrm{F}} \exp\{\mu\frac{Q^{\mathrm{F}}-1}{1-\epsilon Q^{\mathrm{F}}}\},$$
where $N$ is the total number of entries in the histogram.
It is found that such a fit function can describe
large variety of LED spectra for different SiPMs, bias voltages and LED intensities.

Using this formula one can calculate the mean and the second central moment (variance)
of the modelled response function. 
To calculate the mean one should multiply the average number of detected photons $\mu$ 
by the average response to one photon and add pedestal: \mbox{$<\!\!R\!\!> = \mu <\!\!x_Q\!\!> /(1-\epsilon) + 
<\!\!B\!\!>$}. The second central moment can be written as 
$\rm{var}(R) = \int L(x)\,x^2\,dx - \left(\int L(x)\,x\,dx\right)^2\ + \rm{var}(B)$.
The variance of $L$ distribution can be found using derivatives 
of $L^{\mathrm{F}}$: it is proportional to
$- {L^{\mathrm F}}''(0) + {L^{\mathrm F}}'(0)^2 $.
For the normal distribution of $Q$ the calculation gives
$$\rm{var}(R) = \mu\left[ \Delta^2 \frac{1+\epsilon}{(1-\epsilon)^2} + \frac{(\sigma/\Delta)^2}{1-\epsilon}\right]
 + \rm{var}(B) .$$
The first term in the brackets shows how the variance increases with the cross-talk. The influence of
$\sigma$ through the second term is usually much smaller.
Using the {\it measured} SiPM calibration spectrum $R_{\rm data}$ one can calculate 
\mbox{$<\!\!R_{\rm data}\!\!>$} and \mbox{$\rm{var}(R_{\rm data})$} and use
the formulae above to constrain $\mu, \Delta$ and $\epsilon$. 
On the other hand $\Delta$ can be determined from the spectral power distribution $|R_{\rm data}^F|^2$,
where $|R_{\rm data}^F|$ is the absolute value of $R_{\rm data}$ Fourier transform. It has a maximum
at the harmonic with the period $\Delta$. $|R_{\rm data}^F|^2$ can be viewed as a Fourier transform of the
autocorrelation function $\int R(x+y) R(x) \, dx$. Since the statistical fluctuations in different bins 
($x$ and $x+y$) of the SiPM response spectrum are not correlated, their contribution to the
autocorrelation function is small, and they do not produce harmonic in the
$|R_{\rm data}^F|^2$ power spectrum comparable in magnitude with the main harmonic with period $\Delta$. 
This is similar to the Fourier decomposition of the white noise where two signals at different times
($x$ and $x+y$) are not correlated. 
Thus abscissa of the point in the spectral power spectrum where $|R_{\rm data}^F|^2$ reaches the maximum 
allows one to measure $\Delta$. The ordinate can be used to find $\sigma$, since
it is equal to $\exp\left(2\mu(\xi-1)/(1-\epsilon\xi)\right)$ where
$\xi=\exp\left(-2\left( \pi\sigma / \Delta\right)^2\right)$. 
Here the Fourier transform is defined as $R^F(f) = \int R(x)\,\exp^{-2\pi i\,f x}\,dx$, i.e. with the normalization
similar to the Fast Fourier Transform algorithms.
In this way one can determine all 4
parameters $\mu,\ \epsilon,\ \Delta$ and $\sigma$ from $<\!\!R_{\rm data}\!\!>$, $\rm{var}(R_{\rm data})$
and $|R_{\rm data}^F|^2_{\rm max}$ even without fit. Usually this already gives a good accuracy. 
To achieve even better description of the calibration data one needs to make a fit where these numbers can be 
used as initial fit parameters. 



\end{document}